# Monte Carlo methods for the self-avoiding walk

Alan D. Sokal[a]

[a]Department of Physics, New York University, 4 Washington Place, New York, NY 10003, USA

This article is a pedagogical review of Monte Carlo methods for the self-avoiding walk, with emphasis on the extraordinarily efficient algorithms developed over the past decade. Many more details can be found in [1].

## 1. INTRODUCTION

This talk has no direct relation to QCD; it's therefore intended as *entertainment*. My goal is to show a nontrivial statistical-mechanics problem for which we have successfully developed collective-mode algorithms that *completely eliminate* the critical slowing-down. Hopefully some of these ideas can be adapted to problems in random surfaces, and possibly even to spin and gauge theories. Anyone who wants more details (and more references) is referred to my review article [1].

So, what is a self-avoiding walk (SAW)? Obviously, it's a nearest-neighbor path (on some given lattice) that never visits any site more than once. Figure 1 shows some typical $N$-step SAWs with $N = 100, 1000, 10000$ on the square lattice, scaled down by a factor $N^\nu$ with $\nu = \frac{3}{4}$ (that turns out to be the right critical exponent in dimension $d = 2$). Already we can see that a long SAW is going to be a fractal object. Moreover, the shape and size of these SAWs are completely different from the shape and size of ordinary random walks (which have $\nu = \frac{1}{2}$); this is a different universality class. Note, finally, that even by looking closely at the 10000-step SAW one can't tell whether it's strictly self-avoiding or merely self-repelling; nevertheless, the universality class doesn't depend on such short-distance details.

The purpose of this talk is to explain to you how I generated these three SAWs.

But first of all one should ask: Who cares? Why is the SAW an interesting model? The answer is twofold. On the one hand, the SAW is a model of a linear polymer molecule (such as polystyrene) in a good solvent. Furthermore, long chains are experimentally accessible:

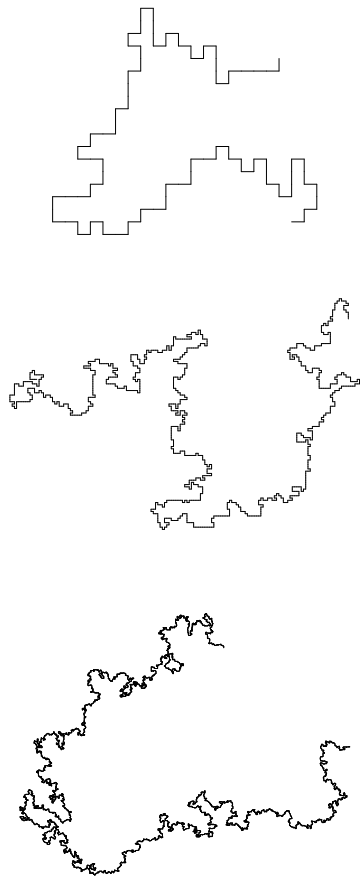

Figure 1. Typical $N$-step SAWs with $N = 100, 1000, 10000$ on the square lattice, scaled down by a factor $N^\nu$ with $\nu = 3/4$. The 10000-step SAW was *not* prepared specially for this conference in Australia.

one can reach $N$ as high as $\sim 10^5$. Secondly, the SAW is isomorphic to the $O(n)$ $\sigma$-model [also known as the $n$-vector model] analytically continued to $n = 0$; more precisely, the SAWs are the strong-coupling diagrams of the $\sigma$-model. So the SAW belongs to a family we already know and love: its siblings include the Ising ($n = 1$), $XY$ ($n = 2$), classical Heisenberg ($n = 3$) and spherical ($n = \infty$) models.

The long-chain limit $N \to \infty$ of the SAW is thus a *critical phenomenon* in the usual sense: it's described by a continuum quantum field theory, so that one expects *universal* behavior with the usual plethora of critical exponents, universal amplitude ratios, universal scaling functions, and so forth. Many of these theoretical predictions are experimentally testable by light scattering on dilute polymer solutions.

In fact, the SAW is an *exceptionally favorable* "laboratory" for the numerical study of critical phenomena, for three reasons:

1) There are no finite-volume effects; one can study directly an $N$-step SAW in an *infinite* lattice.

2) There is no $L^d$ factor in the computational work; one doesn't have to worry about the sites where the walk might be but isn't.

3) There is, to be sure, critical slowing-down; but this can be overcome by cleverly designed collective-mode algorithms.

## 2. DEFINITIONS

Let $c_N$ [resp. $c_N(x)$] be the number of $N$-step SAWs on $\mathbb{Z}^d$ starting at the origin and ending anywhere [resp. ending at $x$]. These quantities are believed to have the asymptotic behavior

$$c_N \sim \mu^N N^{\gamma - 1} \tag{1}$$
$$c_N(x) \sim \mu^N N^{\alpha_{sing} - 2} \quad (x \text{ fixed} \neq 0) \tag{2}$$

as $N \to \infty$. [Compare ordinary random walks, for which $c_N = (2d)^N$ and $c_N(x) \sim (2d)^N N^{-d/2}$.] Here $\mu$ is called the *connective constant* of the lattice: it is like a critical temperature, and therefore lattice-dependent. By contrast, $\gamma$ and $\alpha_{sing}$ are *critical exponents*, and thus expected to be universal among lattices of a given dimension $d$.

Now consider, for each fixed $N$, the probability distribution in which each $N$-step SAW gets equal weight. Then we can ask about the *mean-square end-to-end distance* $\langle R_e^2 \rangle$, the *mean-square radius of gyration* $\langle R_g^2 \rangle$, and so forth: all these measures of the size of the SAW are believed to grow as

$$\langle R_e^2 \rangle_N, \langle R_g^2 \rangle_N \sim N^{2\nu} \tag{3}$$

as $N \to \infty$, where $\nu$ is another (universal) critical exponent. Moreover, the ratio $\langle R_g^2 \rangle / \langle R_e^2 \rangle$ — along with many other dimensionless measures of the shape of a SAW — tends as $N \to \infty$ to a universal limiting value.

More information on the SAW can be found in the excellent book by Madras and Slade [2].

*Remark:* In dimension $d = 2$, some of these universal quantities have been computed exactly by conformal-field-theory and Coulomb-gas techniques: for example, $\nu = \frac{3}{4}$ and $\gamma = \frac{43}{32}$ [3]. It's an open question whether all of them can be so computed; for example, the exact value of the limiting ratio $\langle R_g^2 \rangle / \langle R_e^2 \rangle$ is unknown (to four decimal places it's 0.1403 [4–6]).

Different aspects of the SAW can be probed in four different ensembles:

- Fixed-length, fixed-endpoint ensemble (fixed $N$, fixed $x$)
- Fixed-length, free-endpoint ensemble (fixed $N$, variable $x$)
- Variable-length, fixed-endpoint ensemble (variable $N$, fixed $x$)
- Variable-length, free-endpoint ensemble (variable $N$, variable $x$)

The fixed-length ensembles are best suited for studying the critical exponent $\nu$, while the variable-length ensembles are best suited for studying the connective constant $\mu$ and the critical exponents $\alpha_{sing}$ (fixed-endpoint) or $\gamma$ (free-endpoint). Physically, the free-endpoint ensembles correspond to linear polymers, while the fixed-endpoint ensembles with $|x| = 1$ correspond to ring polymers. All these ensembles give equal weight to all walks of a given length; but the variable-length ensembles have considerable freedom in choosing the relative weights of different chain lengths $N$.

Most of the algorithms I'll discuss here work in the fixed-$N$, free-endpoint ensemble. Indeed, somewhat less progress has been made in eliminating critical slowing-down in the other three ensembles, and many fascinating problems remain open. See [1] for details.

## 3. STATIC ALGORITHMS

I call a Monte Carlo algorithm *static* if it produces a sequence of statistically indepdendent samples: that is, it's of the form "call the subroutine and it returns a random SAW", independent of all previous ones.

The most obvious static technique for generating a random $N$-step SAW is *simple sampling*: just generate a random $N$-step *ordinary* random walk, and reject it if it is not self-avoiding; keep trying until success. It is easy to see that this algorithm produces each $N$-step SAW with equal probability. Of course, to save time we should check the self-avoidance as we go along, and reject the walk as soon as a self-intersection is detected.

The trouble with this algorithm is, of course, the exponentially rapid sample attrition for long walks: the probability of an $N$-step walk being self-avoiding is $c_N/(2d)^N \sim (\mu/2d)^N$. Some improvement can be obtained by modifying the walk-generation process so as to produce only walks without immediate reversals; but the success probability still decays like $(\mu/(2d-1))^N$. One can try higher-order variants of simple sampling, in which walks with loops of length $\leq r$ are automatically absent; but now it's non-trivial to ensure that each $N$-step SAW gets generated with equal probability, and one still has exponential attrition (albeit a weaker one). All in all, this approach seems to be a dead end.

So it's perhaps surprising that there *is* a reasonably efficient static algorithm for generating SAWs: it's called *dimerization*, and it's an implementation of the computer scientists' principle of "divide and conquer". To generate an $N$-step SAW, we generate two independent $(N/2)$-step SAWs ("dimers") and attempt to concatenate them. If the result is self-avoiding, we are done; otherwise, we discard the two walks and start again from scratch. This procedure can now be repeated recursively: to generate each of the $(N/2)$-step SAWs, we generate a pair of $(N/4)$-step SAWs and attempt to concatenate them, and so on. For $N \leq$ some cutoff $N_0$, we stop the recursion and generate the SAWs by some primitive method, such as non-reversal simple sampling. The dimerization algorithm can thus be written recursively as follows:

> **function** dim($N$)
> **if** $N \leq N_0$ **then**
>   $\omega \leftarrow$ nrssamp($N$)
>   **return** $\omega$
> **else**
>   $N_1 \leftarrow \lfloor N/2 \rfloor$ (integer part)
>   $N_2 \leftarrow N - N_1$
>   start:
>     $\omega^{(1)} \leftarrow$ dim($N_1$)
>     $\omega^{(2)} \leftarrow$ dim($N_2$)
>     $\omega \leftarrow \omega^{(1)} \circ \omega^{(2)}$ (concatenation)
>     **if** $\omega$ is not self-avoiding **goto** start
>   **return** $\omega$
> **endif**

It is easy to prove inductively that algorithm dim produces each $N$-step SAW with equal probability, using the fact that the subroutine nrssamp does so. It is crucial here that after a failure we discard *both* walks and start again *from scratch*.

Let's analyze the efficiency of the dimerization algorithm under the scaling hypothesis

$$c_N \approx A\mu^N N^{\gamma-1} \,. \qquad (4)$$

Let $T_N$ be the mean CPU time needed to generate an $N$-step SAW by algorithm dim. Now, the probability that the concatenation of two random $(N/2)$-step SAWs yields an $N$-step SAW is

$$p_N = \frac{c_N}{(c_{N/2})^2} \approx B^{-1} N^{-(\gamma-1)} \,, \qquad (5)$$

where $B = A/4^{\gamma-1}$. We will need to generate, on average, $1/p_N$ pairs of $(N/2)$-step SAWs in order to get a single $N$-step SAW; hence

$$T_N \approx BN^{\gamma-1} 2 T_{N/2} \,. \qquad (6)$$

(We have neglected here the time needed for checking the intersections of the two dimers; this time is linear in $N$, which, as will be seen shortly, is negligible compared to the time $2T_{N/2}$ for generating the two dimers.) Iterating this $k$ times, where $k = \log_2(N/N_0)$ is the number of levels, we obtain

$$T_N \approx C_0 N^{C_1 \log_2 N + C_2} , \qquad (7)$$

where

$$C_1 = (\gamma - 1)/2 \qquad (8)$$
$$C_2 = (5 - 3\gamma)/2 + \log_2 A \qquad (9)$$

and $C_0$ depends on $N_0$. Thus, the growth of $T_N$ is slower than exponential in $N$; but if $\gamma > 1$ (which occurs for $d < 4$) it is faster than any polynomial in $N$. Fortunately, however, the constants $C_1$ and $C_2$ are very small, so that in practice $T_N$ behaves like $N^{\approx 2-3}$ up to $N$ of order several thousand (resp. several million) in $d = 2$ (resp. $d = 3$).

This may be the *only* known subexponential-time static algorithm for a nontrivial statistical-mechanical problem.[1] It's an open question whether there exists a polynomial-time algorithm for generating SAWs in dimension $d \leq 4$.

## 4. DYNAMIC ALGORITHMS

Most of the Monte Carlo algorithms familiar to quantum field theorists are *dynamic* algorithms: that is, they generate a sequence of correlated samples from some Markov process having the desired probability distribution as its unique equilibrium distribution. The main problem with dynamic algorithms, as we all know, is *critical slowing-down*[2]: the *autocorrelation time* $\tau$ of the Monte Carlo dynamics[3] grows as the critical point is approached. For the SAW, "criticality" means $N \to \infty$.

The elementary moves in a SAW Monte Carlo algorithm can be classified according to whether they are

- $N$-conserving or $N$-changing
- endpoint-conserving or endpoint-changing
- local, bilocal or non-local

Obviously, fixed-$N$ algorithms must use only $N$-conserving moves, while variable-$N$ algorithms are free to use both $N$-conserving and $N$-changing moves (and indeed *must* use some of the latter in order to satisfy ergodicity). An analogous statement holds for fixed-$x$ and variable-$x$ algorithms with regard to endpoint-conserving and endpoint-changing moves.

The most important distinction is between local, bilocal and non-local moves. Pure local and bilocal algorithms are easy to invent, but they lead to critical slowing-down (just as for the analogous algorithms in QCD). Non-local algorithms are harder to invent, but they offer at least the *possibility* of radically reducing or even completely eliminating the critical slowing-down.

Obviously there's no space here to review in detail all the known algorithms for the SAW (see [1]). Instead, I'd like to present one algorithm of each type, just to give the flavor of how these algorithms work and how they can be analyzed.

### 4.1. Local algorithms

A *local* move is one that alters only a few consecutive sites ("beads") of the SAW, leaving the other sites unchanged. Otherwise put, a local move excises a small piece from the original SAW and splices in a new local configuration in its place. (Of course, it is always necessary to verify that the proposed new walk is indeed self-avoiding.) Let's restrict attention for simplicity to $N$-*conserving* local moves.

Figure 2 shows all the possible one-bead local moves (on a hypercubic lattice). Move A is a "one-bead flip" (also called "kink-jump"); it is the only one-bead internal move. Moves B and C are end-bond rotations.

Figure 3 shows all the possible *internal* two-bead moves. Move D is a "180° crankshaft". Move E is a "90° crankshaft"; of course it is possible only in dimension $d \geq 3$. Move F is a "two-bead L-flip". Move G permutes three successive mutually perpendicular steps (which lie along the edges of a cube); again this is possible only in di-mension $d \geq 3$. I leave it to the reader to construct the list of two-bead end-group moves.

---

[1] Here "nontrivial" is meant to exclude models like independent percolation, Gaussian fields, etc.
[2] See [7] for an introduction to dynamic Monte Carlo methods and critical slowing-down.
[3] In fact there are *several distinct* autocorrelation times — notably the exponential autocorrelation time $\tau_{exp}$ and the integrated autocorrelation times $\tau_{int,A}$ for various observables $A$ — and (contrary to much belief) these can have *different* dynamic critical exponents $z_{exp}$ and $z_{int,A}$. See [8] for discussion.

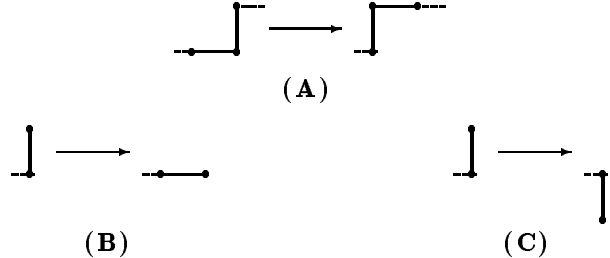

Figure 2. All one-bead local moves. (A) One-bead flip. (B) 90° end-bond rotation. (C) 180° end-bond rotation.

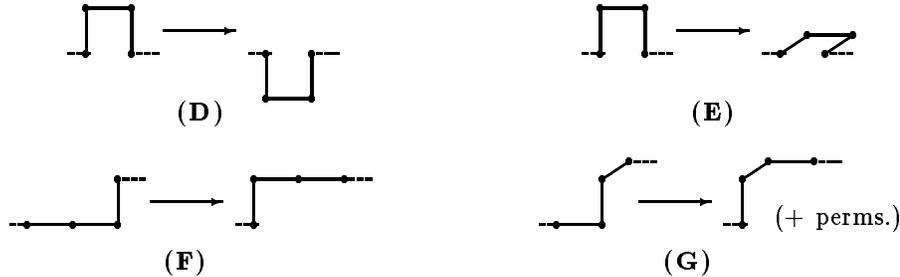

Figure 3. All internal two-bead local moves. (D) 180° crankshaft. (E) 90° crankshaft ($d \geq 3$ only). (F) Two-bead L-flip. (G) Cube permutation ($d \geq 3$ only).

A local algorithm can be built by combining any subset of local moves: for example, one popular algorithm (used heavily by chemical physicists) employs moves A–E. Unfortunately, *all* local $N$-conserving algorithms have a fatal flaw: they are *nonergodic* for sufficiently large $N$.[4] For algorithms based on moves of $k$ or fewer beads, nonergodicity arises in dimension $d = 2$ for all $N \geq 16k + 63$, and for quite a few smaller $N$ as well [9, Theorem 1].

Even if we put aside the problem of nonergodicity, the local algorithms have severe critical slowing-down. A plausible heuristic argument suggests that $\tau \sim N^{2+2\nu} \sim N^{\geq 3}$: Consider the evolution of the center-of-mass vector of the chain. At each elementary move, a few beads of the chain move a distance of order 1; so $\Delta r_{CM} \sim 1/N$. But in order to traverse its equilibrium distribution, $r_{CM}$ must change by something of order its standard deviation, which is $\sim N^{\nu}$. Assuming that $r_{CM}$ executes a random walk, this takes a time $\sim N^{2+2\nu}$. It would be interesting to know whether this prediction for the dynamic critical exponent is exact or merely approximate.[5] In any case, the $N^{\geq 3}$ critical slowing-down of the local algorithms makes it difficult in practice to get beyond $N \sim 100$.

### 4.2. Bilocal algorithms

A *bilocal* move is one that alters two disjoint small groups of consecutive sites (or steps) of the walk; these two groups may in general be very far from each other. Here are some examples:

- The *slithering-snake* (or *reptation*) move, which deletes a bond from one end of the walk and appends a new bond (in an arbitrary direction) at the other end [Figure 4].
- The *kink transport* move, which deletes a kink

---

[4] At least in dimensions $d = 2, 3$. The result is probably true also in dimensions $d \geq 4$, but it has not yet been proven.

[5] Usually dynamic critical exponents are *not* expressible in terms of static ones, except for trivial (Gaussian-like) models.

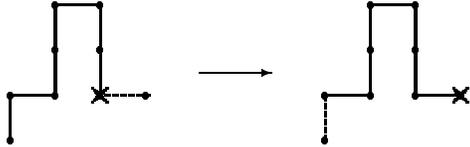

Figure 4. The slithering-snake (reptation) move. Head of the walk is indicated by ×. Dashed lines indicate the proposed new step (resp. the just-abandoned old step).

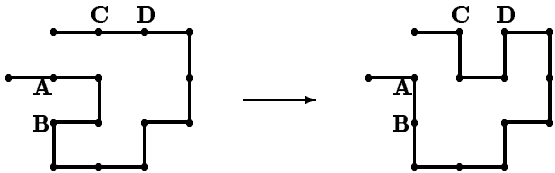

Figure 5. The kink-transport move. A kink has been cleaved from AB and attached at CD. Note that the new kink is permitted to occupy one or both of the sites abandoned by the old kink.

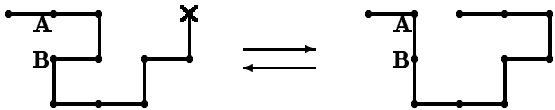

Figure 6. The kink-end reptation ($\longrightarrow$) and end-kink reptation ($\longleftarrow$) moves. In $\longrightarrow$, a kink has been cleaved from AB and two new steps have been attached at the end marked ×. Note that the new end steps are permitted to occupy one or both of the sites abandoned by the kink.

at one location along the walk and inserts a kink (in an arbitrary orientation) at another location [Figure 5].

• The *kink-end reptation* move, which deletes a kink at one location along the walk and appends two new bonds (in arbitrary directions) at one of the ends of the walk [Figure 6 $\longrightarrow$].

• The *end-kink reptation* move, which deletes two bonds from one of the ends of the walk and inserts a kink (in an arbitrary orientation) at some location along the walk [Figure 6 $\longleftarrow$].

Bilocal algorithms can be either nonergodic (e.g. pure reptation[6]) or ergodic (e.g. various combinations). As for the critical slowing-down, a plausible heuristic argument suggests that $\tau \sim N^2$: the SAW transforms itself by random back-and-forth slithering along the chain; after $\sim N^2$ moves, this slithering will have carried it $N$ steps, at which time all the original bonds of the chain will have disappeared and been replaced by random new ones. It would be interesting to know whether this prediction for the dynamic critical exponent is exact or merely approximate. In practice, with a $N^{\approx 2}$ algorithm one can reach $N \sim 1000$.

The study of bilocal algorithms is still in its infancy. We need to understand better the issues of ergodicity and critical slowing-down, and the delineation of dynamic universality classes.

*Remark:* The *BS algorithm* [10] is a *variable-N* algorithm closely related to reptation: it is ergodic and can be proven to have $\tau \sim N^{\approx 2}$.

### 4.3. Non-local algorithms

The possibilities for non-local moves are almost endless, but it is very difficult to find one which is *useful* in a Monte Carlo algorithm. There are two reasons for this: Firstly, since a non-local move is very radical, the proposed new walk usually violates the self-avoidance constraint. (If you move a large number of beads around, it becomes very likely that *somewhere* along the walk two beads will collide.) It is therefore a non-trivial problem to invent a non-local move whose acceptance probability does not go to zero too rapidly as $N \to \infty$. Secondly, a non-local move usually costs a CPU time of order $N$ (or in any case $N^p$ with $p > 0$), in contrast to order 1 for a local or bilocal move. It is non-trivial to find moves whose effects justify this expenditure (by reducing $\tau_{int,A}$ more than they increase $T_{CPU}$).

One extremely successful non-local algorithm is the *pivot algorithm* [11–13]. Here we choose some site along the walk as a pivot point, and apply some symmetry operation of the lattice (e.g. rotation or reflection) to the part of the walk subsequent to the pivot point [Figure 7]. It's easy to prove detailed balance, and with some work ergodicity can be proven as well [13].

---

[6] Exercise for the reader: Find a frozen configuration. One solution can be found in [1, Section 6.4.2].

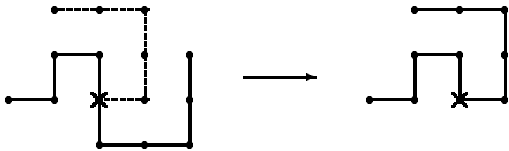

Figure 7. The pivot move (here a $+90°$ rotation). The pivot site is indicated with an ×. Dashed lines indicate the proposed new segment.

At first thought this seems to be a terrible algorithm[7]: for $N$ large, nearly all the proposed moves will get rejected. In fact, this latter statement is true, but the hasty conclusion drawn from it is radically false! The acceptance fraction $f$ does indeed go to zero as $N \to \infty$, roughly like $N^{-p}$; empirically, it is found that the exponent $p$ is $\approx 0.19$ in $d = 2$ and $\approx 0.11$ in $d = 3$. But this means that roughly once every $N^p$ moves one gets an acceptance. And the pivot moves are very radical: one might surmise that after very few accepted moves (say, 5 or 10) the SAW will have reached an "essentially new" configuration. One conjectures, therefore, that the autocorrelation time $\tau$ of the pivot algorithm behaves as $\sim N^p$.[8] On the other hand, a careful analysis of the computational complexity of the pivot algorithm shows that one accepted move can be produced in a computer time of order $N$.[9] Combining these two facts, we conclude that one "effectively independent" sample (at least as regards *global* observables) can be produced in a computer time of order $N$. This is vastly better than the $N^{2+2\nu}$ of the local $N$-conserving algorithms and the $N^{\approx 2}$ of the bilocal algorithms. Indeed, this order of efficiency cannot be surpassed by any algorithm which computes each site on successive SAWs, for it takes a time of order $N$ simply to *write down* an $N$-step walk!

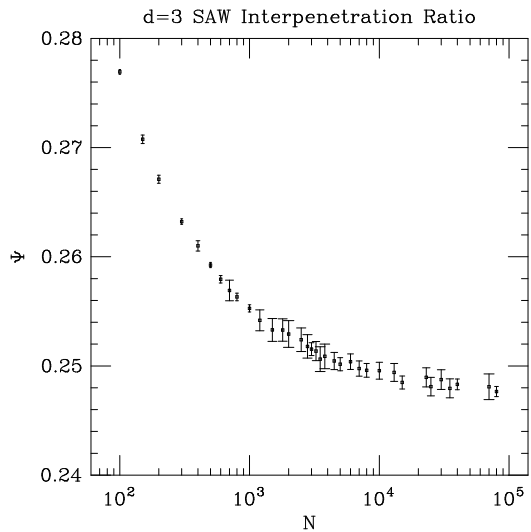

Figure 8. Interpenetration ratio $\Psi$ versus chain length $N$, for SAWs in $d = 3$. Error bar is one standard deviation. Data from [5].

In practice, with the pivot algorithm one can reach chain lengths $N$ of order $10^5$, with high statistics [5]. One can then obtain extremely high-precision ($\pm 0.001$) estimates of critical exponents and universal amplitude ratios, with good control over *corrections to scaling*.[10] An example is shown in Figure 8, which plots the *interpenetration ratio* $\Psi$ for pairs of $N$-step SAWs (it's a kind of dimensionless renormalized coupling constant) as a function of $N$. We can estimate the limiting value $\Psi^* = 0.2471 \pm 0.0003$, and we can see clearly the strong corrections to scaling. This plot also shows that most of the last 40 years of polymer theory is *wrong* (albeit fixable), but that's another story [14,5].

---

[7] Indeed, this was my initial view. See [13, footnote 3] for a *mea culpa*.

[8] Things are in fact somewhat more subtle: heuristic arguments and numerical evidence suggest that $\tau_{int,A} \sim N^p$ for observables $A$ that are "essentially global" (like $R_e^2$ and $R_g^2$), while $\tau_{int,A} \sim N^{1+p}$ for "local" observables (like the number of $90°$ angles in the walk). See [13, Sections 3.2, 3.3, 4.2 and 4.3] for details.

[9] One should use a hash table to test self-avoidance, and look for a self-intersection starting at the pivot point and working outwards. Failed moves can then be detected in a mean CPU time of order $N^{1-p}$ (because the self-intersection is most likely to be near the pivot point), while successes of course require a time of order $N$. See [13, Sections 3.4 and 4.4] for details.

[10] Corrections to scaling are a very serious problem when one is aiming for this level of accuracy. For example, in the $d = 3$ SAW, to get the exponent $\nu$ with a systematic error of less than 0.001 from a pure power-law fit $\langle R_g^2 \rangle \sim N^{2\nu}$, one must *throw away* all the walks with $N \lesssim 1000$ [5]!

### 4.4. Other algorithms

Well, there are many more interesting algorithms that I don't have time to discuss. For the variable-$N$, free-endpoint ensemble we have the *slithering-tortoise (BS)* [10] and the *join-and-cut* [15] algorithms. For the variable-$N$, fixed-endpoint ensemble we have the *BFACF* [16] and *BFACF + cut-and-paste* [17] algorithms. But there's lots of room for inventing new and better non-local algorithms for these ensembles. And for *all* the algorithms described here, we need to understand better their dynamic critical behavior.

A big open problem is to devise good non-local algorithms for the SAW with *nearest-neighbor self-attraction*, which has a collapse transition (called the theta point) as the temperature is varied. The pivot algorithm (for example) can easily be modified to handle a nearest-neighbor interaction, by inserting a Metropolis accept/reject step; but its efficiency deteriorates markedly in the neighborhood of the theta point and even more drastically in the collapsed regime.

Another open problem is to generalize these algorithms to related models such as branched polymers, random surfaces, etc. For these problems local algorithms are known, but good non-local algorithms are just beginning to be invented.

## 5. CONCLUSIONS

The SAW is an extraordinary success story for the development of non-local (collective-mode) algorithms: the CPU time needed to generate one "effectively independent" SAW has been reduced from order $e^{\lambda N}$ (simple sampling) to $N^{C_1 \log N + C_2}$ (dimerization) to $N^{\geq 3}$ (local algorithms) to $N^{\approx 2}$ (bilocal algorithms) to $N$ (pivot algorithm).

Here's the recipe for success of a non-local algorithm: Offer the system a collective-mode move that *it wants*. (The move has to be *radical* but *sensible*.) And do it in a CPU time of order $N$ ($\simeq$ volume in a spin or gauge system) or in any case not much more. Of course, this is not really a "recipe"; it's more like "guidelines". To invent a good non-local algorithm, one needs *physical insight* combined with *cleverness*.

The big open question is: Can any of these algorithms from polymer physics inspire successful collective-mode algorithms for spin or gauge theories? Or vice versa?

Oh, yes, I almost forgot: the three SAWs in Figure 1 were produced using the pivot algorithm.

This research was supported by the Petroleum Research Fund and by NSF grant DMS-9200719.